\documentclass[12pt]{article}
\usepackage[a4paper, total={7in, 10in}]{geometry}
\usepackage[parfill]{parskip}
\usepackage{physics, tensor, float, subcaption}
\usepackage{graphicx}
\graphicspath{ {Plots/} }
\usepackage{jhep-mod}
\usepackage{bm}
\usepackage{soul}
\usepackage{amssymb,amsmath,amsthm}
\usepackage{mathrsfs}
\usepackage[utf8]{inputenc}
\usepackage{enumerate}
\usepackage{bigints}
\usepackage{xcolor}
\usepackage{appendix}
\usepackage{graphicx}
\usepackage{float}
\usepackage{tikz}
\usepackage{setspace}
\usepackage{cancel}
\usepackage{doi}
\definecolor{purple}{rgb}{1,0,1}
\definecolor{lime}{HTML}{A6CE39}
\newcommand{\orcidicon}{%
	\begin{tikzpicture}
	\draw[lime, fill=lime] (0,0) 
		circle [radius=0.16] 
		node[white] {{\fontfamily{qag}\selectfont \tiny ID}};
	\draw[white, fill=white] (-0.0625,0.095) 
		circle [radius=0.007];
	\end{tikzpicture}
	\hspace{-5mm}
}
\newcommand\orcidJosh{{\href{https://orcid.org/0000-0003-1200-7261}{\orcidicon}}}
\newcommand\orcidThomas{{\href{https://orcid.org/0000-0002-0314-4136}{\orcidicon}}}
\newcommand\orcidAlex{{\href{https://orcid.org/0000-0002-1763-3563}{\orcidicon}}}
\newcommand\orcidMatt{{\href{https://orcid.org/0000-0003-1088-6485}{\orcidicon}}}
\begin{document}
%=================================
%=================================
%=================================

\title{\vspace{-25pt}\huge{
Constant-$r$ geodesics in the\\
 Painlev\'e--Gullstrand form of \\ 
Lense--Thirring spacetime
}}

%=================================
%=================================
%=================================
\author{
\Large
Joshua Baines\!\orcidJosh$^1$, Thomas Berry\!\orcidThomas$^2$, 
\\
Alex Simpson\!\orcidAlex$^1$,\! 
{\sf  and} Matt Visser\!\orcidMatt$^1$}

%=================================
%=================================
%=================================
%=================================
\affiliation{
$^1$ School of Mathematics and Statistics, Victoria University of Wellington, 
\\
\null\qquad PO Box 600, Wellington 6140, New Zealand.}
\affiliation{
$^2$ Robinson Institute,
Victoria University of Wellington, 
\\
\null\qquad PO Box 600, Wellington 6140, New Zealand.}
%=================================
%=================================
\emailAdd{joshua.baines@sms.vuw.ac.nz}
\emailAdd{thomas.berry@vuw.ac.nz}
\emailAdd{alex.simpson@sms.vuw.ac.nz}
\emailAdd{matt.visser@sms.vuw.ac.nz}
%=================================
%=================================

\abstract{
\vspace{1em}

Herein we explore the \emph{non-equatorial {constant-$r$}} ({``quasi-circular''}) geodesics  (both timelike and null) in the Painlev\'{e}--\-Gullstrand variant of the Lense--Thirring spacetime recently introduced by the current authors. 
Even though the spacetime is not spherically symmetric, shells of {constant-$r$} geodesics still exist.
Whereas the radial motion is (by construction) utterly trivial, determining the allowed \emph{locations} of these {constant-$r$} geodesics is decidedly non-trivial, and the stability analysis is equally tricky.
Regarding the angular motion, these {constant-$r$} orbits will be seen to exhibit  both precession and nutation --- typically with incommensurate frequencies. Thus {this constant-$r$} geodesic motion, though \emph{integrable} in the precise technical sense, is generically surface-filling, with the orbits completely covering a symmetric equatorial band which is a segment of a 
{spherical surface}, {(a so-called  ``spherical zone'')}, and whose {latitudinal} extent is governed by delicate interplay between the orbital angular momentum and the Carter constant.
The situation is qualitatively similar to that for the (exact) Kerr spacetime --- but we now see that any physical model having the same slow-rotation weak-field limit as general relativity will still possess non-equatorial {constant-$r$}  geodesics. 

\bigskip
\noindent
{\sc Date:} Friday 18 February  2022; Thursday 3 March 2022;  21 June 20222; \\
19 July 2022;  \LaTeX-ed \today


\bigskip
\noindent{\sc PhySH:} 
Gravitation
}

%=================================
\maketitle
%=================================
\def\tr{{\mathrm{tr}}}
\def\diag{{\mathrm{diag}}}
\def\cof{{\mathrm{cof}}}
\def\pdet{{\mathrm{pdet}}}
\def\d{{\mathrm{d}}}
\def\K{{\mathcal{K}}}
\def\O{{\mathcal{O}}}
\parindent0pt
\parskip7pt
\newcommand{\C}{\mathcal{C}}
\def\sign{{\mathrm{sign}}}
\def\floor{{\mathrm{floor}}}
\def\Frac{{\mathrm{frac}}}
%===================================
\section{Introduction}
\label{S:intro}
%===================================

The Kerr spacetime~\cite{Kerr,Kerr-Texas,kerr-newman, kerr-intro, kerr-book, kerr-book-2}, perhaps the pre-eminent exact solution of the Einstein equations of vacuum general relativity, is both a standard textbook exemplar~\cite{MTW,Wald,Weinberg,Adler-Bazin-Schiffer,  Hobson, D'Inverno,Hartle,Carroll},\break
and is increasingly of central importance to both observational  and theoretical \break
astrophysics~\cite{Berti:2015,Yunes:2016,Cardoso:2019,Barack:2006, Bambi:2019, Barack:2018,LISA:2020}. 
One key issue of particular importance  is a full understanding of the geodesics --- and the fact that despite the lack of spherical symmetry (the Kerr spacetime is merely stationary and axisymmetric, so that the Birkhoff theorem does not apply~\cite{Birkhoff, Jebsen, Deser, Ravndal, Skakala}), there are still a multitude of  {constant-$r$} ``quasi-circular'' geodesics which are \emph{not} confined to the equatorial plane.  (Contrast, for example, the discussion in~\cite{Edery:2006} with  that of~\cite{Hod:2011,Warburton:2013,Hod:2012,Teo:2020,Tavlayan:2020}.)
{Sometimes these constant-$r$ ``quasi-circular'' geodesics are referred to as ``spherical geodesics''.}

\clearpage

It should be noted that the non-equatorial {constant-$r$} null geodesics are particularly important tools for studying photon rings and black hole silhouettes~\cite{Broderick:2013, Johannsen:2016, Broderick:2008,  Pappas:2018, Gralla:2019, Perlick:2021}.
Similarly non-equatorial {constant-$r$} timelike geodesics are particularly important tools for studying off-axis accretion disks and their related ISCOs and OSCOs~\cite{Bambi:2017, Vincent:2020, Chael:2021, Berry:ISCOs,Boonserm:2019}.

\begin{figure}[!htbp]
\begin{center}
\includegraphics{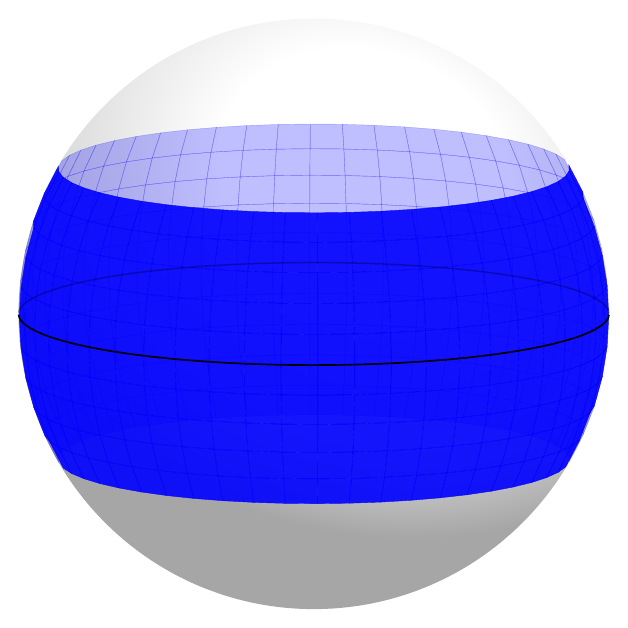}
\caption{Schematic depiction of the generic situation, where the constant-$r$ geodesics have incommensurate azimuthal and declination frequencies, and so sweep out a surface-filling symmetric equatorial band, (a spherical zone). The width of the equatorial band is controlled by a delicate interplay between the Carter constant and the azimuthal angular momentum.}
\label{F:band}
\end{center}
\end{figure}

In the current article we shall be interested is seeing how much of this qualitative structure survives once one moves away from the exact Kerr spacetime, specifically once one considers the Painlev\'e--Gullstrand  version of the weak-field slow-rotation Lense--Thirring spacetime. 
The weak-field slow-rotation Lense--Thirring spacetime was originally introduced in 1918~\cite{Lense-Thirring,Pfister}, while
the current authors have recently introduced, and extensively explored, a novel Painlev\'e--Gullstrand variant~\cite{painleve1,painleve2, gullstrand, poisson, Faraoni:2020, Boonserm:2017} of the Lense--Thirring spacetime~\cite{PGLT1,PGLT2,PGLT3}. 
{We shall soon see that the generic situation is as pictured in figure~\ref{F:band}.}
{The key physical reason underpinning the existence of these {constant-$r$} geodesics comes from the fact that the Kerr, Schwarzschild, and Painlev\'e--Gullstrand--Lense--Thirring spacetimes all possess a non-trivial Killing tensor and associated Carter constant.}

%=========================================
%=========================================
%=========================================
%=========================================

\clearpage
%===================================
\section{Basic framework}
\label{S:basics}
%===================================

The line-element of interest is~\cite{PGLT1,PGLT2,PGLT3}: 
\begin{equation}
\label{E:pglt}
\d s^2 = - \d t^2 +\left\{\d r+\sqrt{\frac{2m}{r}} \; \d t\right\}^2
+ r^2 \left\{\d\theta^2+\sin^2\theta\; \left(\d\phi - {2J\over r^3} \d t\right)^2\right\} \ .
\end{equation}
This line-element is somewhat related to the ``river'' model for black holes~\cite{river} --- it exhibits both unit lapse~\cite{Unit-lapse}, and flat spatial $3$-slices~\cite{PGLT1,PGLT2,PGLT3} --- the presence of flat spatial $3$-slices being incompatible with the exact Kerr spacetime~\cite{Valiente-Kroon:2004a, Valiente-Kroon:2004b, Jaramillo:2007, Kerr-Darboux}.
Furthermore this line element possesses a non-trivial Killing tensor~\cite{PGLT2,PGLT3}:
\begin{equation}
K_{ab} \; \d x^a \; \d x^b = 
r^4\left\{\d\theta^2 + \sin^2\theta \left(\d\phi - {2J\over r^3} \d t\right)^2 \right\}.
\end{equation}
This Killing tensor was found by applying the algorithm presented in~\cite{Papadopoulos:2018, Papadopoulos:2020,Benenti:1979}. Once found, one can easily verify that $\nabla_{(c}K_{ab)}= K_{(ab;c)} = 0$.
For any affine parameter $\lambda$, the (generalized) Carter constant is then~\cite{PGLT2,PGLT3}:
\begin{equation} 
\label{E:Carter0}
\mathcal{C}=K_{ab}\;\frac{\d x^a}{\d\lambda}\,\frac{\d x^b}{\d\lambda}=r^4\left[ \left(\frac{\d\theta}{\d\lambda}\right)^2 + \sin^2\theta \left(\frac{\d\phi}{\d\lambda}-\frac{2J}{r^3}\frac{\d t}{\d\lambda}\right)^2 \right] \ .
\end{equation}
 By construction $\C \geq 0$.
 Without loss of generality we choose $\lambda$ to be future-directed, $\d t/\d\lambda > 0$.
%

%\clearpage

In addition to the Carter constant,  we have three other conserved quantities. Two (the energy and azimuthal component of angular momentum) come from the time-translation and axial Killing vectors~\cite{PGLT2,PGLT3}:%\enlargethispage{20pt}
\begin{equation}\label{E}
E 
= \left(1-\frac{2m}{r}-\frac{4J^2\sin^2\theta}{r^4} \right)\frac{\d t}{\d\lambda}-\sqrt{\frac{2m}{r}}\frac{\d r}{\d\lambda}+\frac{2J\sin^2\theta}{r}\frac{\d\phi}{\d\lambda} \ ;
\end{equation}
\begin{equation} \label{L}
L =  
r^2\sin^2\theta\frac{\d\phi}{\d\lambda} -\frac{2J\sin^2\theta}{r}\frac{\d t}{\d\lambda} \ .
\end{equation}

The final conserved quantity, the ``mass-shell constraint'', 
$\epsilon \in \{0,-1\}$ for null and timelike geodesics respectively, 
comes from the trivial Killing tensor (the metric):
\begin{equation}\label{Geo_eqn}
\begin{split}
\epsilon=g_{ab}\frac{\d x^a}{\d\lambda}\frac{\d x^b}{\d\lambda}= & -\left(\frac{\d t}{\d\lambda}\right)^2+\left(\frac{\d r}{\d\lambda}+\sqrt{\frac{2m}{r}}\frac{\d t}{\d\lambda}\right)^2\\
& + r^2\left[\left(\frac{\d\theta}{\d\lambda}\right)^2+\sin^2\theta\left(\frac{\d\phi}{\d\lambda}-\frac{2J}{r^3}\frac{\d t}{\d\lambda}\right)^2\right] \ .
\end{split}
\end{equation}

\clearpage
Simplify these four conserved quantities by re-writing them as follows~\cite{PGLT2,PGLT3}:
\begin{equation} \label{L_2}
L=r^2\sin^2\theta \left(\frac{\d\phi}{\d\lambda}-\frac{2J}{r^3}\frac{\d t}{\d\lambda}\right) \ ;
\end{equation}
\begin{equation} \label{C_2}
\mathcal{C}=r^4\left(\frac{\d\theta}{\d\lambda}\right)^2 +{L^2\over \sin^2\theta} \ ;
\end{equation}
\begin{equation}\label{Geo_eqn_2}
\epsilon=  -\left(\frac{\d t}{\d\lambda}\right)^2+\left(\frac{\d r}{\d\lambda}+\sqrt{\frac{2m}{r}}\frac{\d t}{\d\lambda}\right)^2 +{\mathcal{C}\over r^2} \ ;
\end{equation}
\begin{equation}\label{E_2}
E=\left(1-\frac{2m}{r}\right)\frac{\d t}{\d\lambda} - \sqrt{\frac{2m}{r}}\frac{\d r}{\d\lambda} + \frac{2J}{r^3}L \ .
\end{equation}
In particular $L^2 \leq \C$. 
For (generic) geodesic trajectories we have~\cite{PGLT2,PGLT3}:
\begin{equation}
\frac{\d r}{\d\lambda}= S_r \sqrt{X(r)} \ ;
\label{E:r(lambda)}
\end{equation}
\begin{equation}
\frac{\d t}{\d\lambda}= \frac{E-2JL/r^3+ S_r \sqrt{(2m/r)X(r)}}{(1-2m/r)} \ ;
\label{E:t}
\end{equation}
\begin{equation}
\frac{\d\theta}{\d\lambda}= S_{\theta}\frac{\sqrt{\mathcal{C}-L^2/\sin^2\theta}}{r^2} \ ,
\label{E:theta}
\end{equation}
\begin{equation}
\frac{\d\phi}{\d\lambda}= \frac{L}{r^2\sin^2\theta} + 2J\; \frac{E-2JL/r^3 + S_{\phi}\sqrt{(2m/r)X(r)}}{r^3(1-2m/r)} \ . \label{E:phi}
\end{equation}

%\clearpage
Here
\begin{eqnarray}
    S_r  &=& \left\{
    \begin{array}{rl}
    +1 & \qquad\mbox{outgoing geodesic} \\
     -1 & \qquad\mbox{ingoing geodesic}
    \end{array}\right. \ ; \label{Sphi}\\
    && \nonumber \\
    S_{\theta} &=& \left\{
    \begin{array}{rl}
    +1 & \qquad\mbox{increasing declination geodesic} \\
     -1 & \qquad\mbox{decreasing declination geodesic}
    \end{array}\right. \ ; \\
    && \nonumber \\
    S_{\phi} &=& \left\{
    \begin{array}{rl}
    +1 & \qquad\mbox{prograde geodesic} \\
     -1 & \qquad\mbox{retrograde geodesic}
    \end{array}\right. \ .
\end{eqnarray}\enlargethispage{20pt}
Furthermore $X(r)$ is explicitly given by the sextic Laurent polynomial:
\begin{equation} \label{X1}
X(r) =\left(E - \frac{2JL}{r^3}\right)^2-\left(1-\frac{2m}{r}\right)\left(-\epsilon + \frac{\mathcal{C}}{r^2}\right) \ ;
\qquad 
\lim_{r\to\infty} X(r) = E^2+\epsilon \ .
\end{equation} \enlargethispage{40pt}
In terms of the roots of this polynomial we can in the generic case write
\begin{equation}
X(r) = {E^2+\epsilon\over r^6} \;\prod_{i=1}^6 (r-r_i) \ .
\end{equation}
We shall now restrict attention to the {constant-$r$}  orbits, $r\to r_0$.

\clearpage
%===================================
\section{{Location of possible constant-$r$ geodesics}}
%===================================
\label{S:circular}
%===================================

First let us analyze the  \emph{lack} of radial motion; this is not entirely  trivial. 

%----------------------------------------------------------------
\subsection{Generalities} 
%----------------------------------------------------------------

Fix our $r$ coordinate to take some fixed value $r=r_0$. 
Hence, since $\d r/\d\lambda = S_r \sqrt{X(r)}$, we must have $X(r_0)=0$. Furthermore,
using the chain rule and the fact that $S_r^2=+1$, we have
\begin{equation}
{\d^2r\over\d \lambda^2}= S_r\,{\d\sqrt{X(r)}\over\d \lambda}  = {1\over2}\;   X'(r) \ .
\end{equation}
So to \emph{remain} at $r_0$ we must also have $X'(r_0)=0$. 
The two conditions 
\begin{equation}
X(r_0)=0 \qquad\hbox{and} \qquad X'(r_0)=0
\end{equation}
imply that $r_0$ is a repeated root of $X(r)$.
The existence of a repeated root will put \emph{some} constraint on the four geodesic constants $E$, $L$, $\epsilon$, and $\C$, (and the spacetime parameters $m$ and $J$); they cannot all be functionally independent.

%\clearpage
Higher derivatives do not lead to extra constraints, since
\begin{equation}
{\d^3r\over\d \lambda^3}
= S_r\,{1\over2}  X''(r) \sqrt{X(r)} \ ;
\end{equation}
\begin{equation}
{\d^4r\over\d \lambda^4}
= \left\lbrace{1\over2}  X'''(r) X(r) + {1\over 4} X''(r) X'(r)\right\rbrace \ ,
\end{equation}
and one sees inductively that all terms in all higher-order derivatives contain either $\sqrt{X(r)}$ or $X'(r)$ as a factor; quantities which we have already seen vanish at $r\to r_0$. Finally we note that \emph{stability} of the {constant-$r$}  orbit is determined by considering 
\begin{equation}
{\d\over\d r} \left({\d^2r\over\d \lambda^2}\right) 
= {1\over2}  \; X''(r) \ .
\end{equation}
Thence if $X''(r_0) >0$ the {constant-$r$}  orbit is \emph{unstable}, if $X''(r_0) =0$ the {constant-$r$}  orbit is \emph{marginal}, and if $X''(r_0) <0$ the {constant-$r$}  orbit is \emph{stable}.
So we are interested in evaluating $\sign\left(X''(r_0)\right)$.
Let us now see what more we can say about the radial location of possible {constant-$r$} (``quasi-circular'') orbits. 

\clearpage
%----------------------------------------------------------------
\subsection{{Constant-$r$ null geodesics}} 
%----------------------------------------------------------------

For massless particles following null geodesics we have $\epsilon\to 0$, and without any loss of generality we can set $E\to 1$. 
That implies that we can write
\begin{eqnarray} 
\label{E:X(r)n}
X(r) &=&\left(1 - \frac{2JL}{r^3}\right)^2-\left(1-\frac{2m}{r}\right)\frac{\mathcal{\C}}{r^2} 
\\[3pt]
&=&
{r^6 - \C r^4 +2(\C m-2 JL) r^3 + 4 J^2 L^2\over r^6}\ ,
\end{eqnarray}
while
\begin{equation} 
\label{E:dX(r)n}
X'(r) =2\; \frac{\C r^4  - 3(\C m-2JL)r^3 - 12 J^2 L^2 }{r^7} \ ,
\end{equation}
and
\begin{equation} \label{E:ddX(r)n}
X''(r) =6\; \frac{-\C r^4  + 4(\C m-2JL) r^3+28 J^2 L^2}{r^8} \ .
\end{equation}
Thence we are interested in {simultaneously} solving
\begin{equation} 
\label{E:X(r0)n}
r_0^6 - \C r_0^4 +2(\C m-2 JL) r_0^3 + 4 J^2 L^2 =0\ ,
\end{equation}
\begin{equation} 
\label{E:dX(r0)n}
\C r_0^4  - 3(\C m-2JL)r_0^3-12 J^2 L^2 =0\ ,
\end{equation}
and evaluating the sign of $X''(r_0)$:
\begin{equation} 
\label{E:ddX(r0)n}
\sign\left({X''(r_0)}\right) = \sign\left({ -\C r_0^4 + 4(\C m-2JL) r_0^3 + 28 J^2 L^2}\right)\ .
\end{equation}
The non-negativity of $\C\geq0$, applied to $X(r_0)=0$, from equation (\ref{E:X(r)n}) immediately implies that $r_0 \geq 2m$.
We also recall that $L^2 \leq \C$. 
The four quantities $\C$, $m$, $JL$, and $r_0$, are subject to two constraints, so only two of these four quantities are functionally independent. More on this point below. 

Since we are interested in the sign of $X''(r_0)$ at a location where $X'(r_0)=0$, we can use that extra information to deduce
\begin{equation} \label{E:ddX(r0)2}
\sign\left(X''(r_0)\right) = \sign\left(\C r_0^4 + 36 J^2 L^2\right) = +1 \ .
\end{equation}
Thus there are no stable {constant-$r$}  null geodesics. (And, as we shall soon see, there are no marginal {constant-$r$} null geodesics either, all of the {constant-$r$} null geodesics are unstable.)

\newpage
Let us now consider several special case solutions to the radial part of the {constant-$r$} null geodesic conditions, $X(r_0)=0=X'(r_0)$:
%\clearpage
\enlargethispage{40pt}
\begin{enumerate}
\item[(i)]
If $JL=0$, corresponding \emph{either} to a non-rotating source, \emph{or} to a zero angular momentum geodesic (ZAMO), then one has the unique unstable {constant-$r$} null geodesic:
\begin{equation}
r_0 = 3m; \qquad \C = 27 m^2; \qquad X''(r_0) = {2\over 3 m^2} =  {6\over r_0^2} > 0. 
\end{equation}
This is the situation familiar from Schwarzschild spacetime; an unstable photon orbit at $r=3m$. 
\item[(ii)] 
If $\C=0$, then $L=0$, and there are no {constant-$r$} null orbits.
\item[(iii)]
If $r_0=2m$ then this implies $\C=0$. This is a sub-case of (ii) above.
\item[(iv)]
If $r_0=3m$ then: either $JL=0$ which is a sub-case of (i) above,
or $\C=0$ which is a sub-case of (ii) above.
\item[(v)]
If $r_0= \sqrt[3]{2JL}\neq 0$ then $\C<0$, which is non-viable, and there are no {constant-$r$}  null orbits.
\item[(vi)]
The generic case is $JL\neq0$, $\C>0$ and $2m < r_0 \not\in \{3m, \sqrt[3]{2JL}\}$.
\end{enumerate}
Now let us consider the generic case:\\
Treat $m$ and $r_0$ as the two independent variables; then we can explicitly solve for $\C(m,r_0)$ and $2JL(m,r_0)$. 
Let us proceed as follows:
If $JL\neq0$ then first solve $X'(r_0)=0$ to find $\C(JL,m, r_0)$. We find
\begin{equation}
\C(JL,m, r_0) = {6(r_0^3-2JL) JL\over r_0^3 (3m-r_0)} \neq 0\ .
\end{equation}
Using this value of $\C(JL,m, r_0)$, solve $X(r_0)=0$ for $2JL(m, r_0)$:
\begin{equation}
2JL(m, r_0) = -{r_0^3\over2}\, {(r_0-3m)\over (2r_0-3m)} .
\end{equation}

Third, substitute these values of $JL(m, r_0)$ back into $\C(JL,m, r_0)$ to yield $\C(m, r_0)$:
\begin{equation}
\C(m, r_0) =9r_0^3\;{(r_0-2m)\over (2r_0-3m)^2} \ .
\end{equation}
Since we must always have $\C> 0$ this limits the generic {constant-$r$}  photon orbits to the range 
$r_0 \in (2m,\infty)$.

\clearpage
Finally, inserting this back into $X''(r_0)$ we see:
\begin{equation}
X''(r_0) = {18\over r_0^2} \; {2(r_0-2m)^2+m^2\over (2r_0-3m)^2} >0.
\end{equation}
Since $X''(r_0)>0$, we again see that all of these {constant-$r$} photon orbits are unstable. 
That is, instead of just having one unstable photon orbit at $r=3m$, once we allow $JL\neq0$ we can arrange unstable photon orbits at arbitrary $r_0 \in (2m,\infty)$.

%----------------------------------------------------------------
\subsection{{Constant-$r$} timelike geodesics}
%----------------------------------------------------------------

For massive particles following timelike geodesics $\epsilon\to -1$, and $E$ is unconstrained. 
That implies that we can write
\begin{equation} \label{E:X(r)t}
X(r) =\left(E - \frac{2JL}{r^3}\right)^2-\left(1-\frac{2m}{r}\right)\left(1+\frac{\mathcal{\C}}{r^2}\right) \ ;
\end{equation}
\begin{equation} \label{E:dX(r)t}
X'(r) ={12JL\over r^4}\left(E - \frac{2JL}{r^3}\right) + {2\C(r-3m)\over r^4}  -{2m\over r^2};
\end{equation}
\begin{equation} \label{E:ddX(r)t}
X''(r) = -{24JL\over r^5}\left(2E-\frac{7JL}{r^3}\right)  - {6\C(r-4m)\over r^5} +{4m\over r^3}.
\end{equation}
Rewrite this as 
\begin{equation} \label{E:X(r)tr}
X(r) ={(E^2-1)r^6 +2mr^5 - \C r^4 +2(\C m -2EJL) r^3+4J^2L^2
\over r^6} \ ;
\end{equation}
\begin{equation} \label{E:dX(r)tr}
X'(r) =- 2 \; {mr^5 -\C r^4 +3(\C m- 2 EJL) r^3 +12 J^2 L^2 \over r^7};
\end{equation}
\begin{equation} \label{E:ddX(r)tr}
X''(r) = 2 \; {2mr^5 -3\C r^4 +12(\C m- 2 EJL) r^3 +84 J^2 L^2\over r^8}.
\end{equation}
As before we are interested in solving $X(r_0)=0=X'(r_0)$, and determining the sign of $X''(r_0)$. 
So we are interested in studying
\begin{equation} 
\label{E:X(r)tn}
(E^2-1)r^6 +2mr^5 - \C r^4 +2(\C m -2EJL) r^3+4J^2L^2=0;
\end{equation}
\begin{equation} 
\label{E:dX(r)tn}
mr^5 -\C r^4 +3(\C m- 2 EJL) r^3 +12 J^2 L^2 =0;
\end{equation}
\begin{equation} 
\label{E:ddX(r)tn}
\sign\{2mr^5 -3\C r^4 +12(\C m- 2 EJL) r^3 +84 J^2 L^2\}.
\end{equation}
The five quantities $E$, $\C$, $m$, $JL$, and $r_0$ are subject to two constraints, so only three of these quantities can be functionally independent.
The positivity of $(1+\C/r^2)>0$, applied to $X(r_0)=0$, immediately implies $r_0 \geq 2m$.
There are several ways of proceeding. 

\clearpage
Let us now consider several special case solutions to $X(r_0)=0=X'(r_0)$:

%\clearpage
\begin{enumerate}\enlargethispage{20pt}
\item[(i)]
If $JL=0$, corresponding \emph{either} to a non-rotating source, \emph{or} to a zero angular momentum geodesic (ZAMO), then for {constant-$r$}  orbits one has:
\begin{equation}
\C = {mr_0^2\over r_0-3m} ; \qquad
E^2 = {(r_0-2m)^2\over r_0(r_0-3m)};
\end{equation}
and
\begin{equation}
X''(r_0) = -{2m\over r_0^3} \; {r_0-6m\over r_0-3m}. 
\end{equation}
Positivity of $\C$ and/or $E^2$ implies $r_0\geq 3m$, and $X''(r_0)$ changes sign at $r_0=6m$. 
This is the situation familiar from Schwarzschild spacetime; an ISCO at $r=6m$, stable orbits for $r_0\in(6m,\infty)$, and unstable orbits for $r_0\in(3m,6m)$. 
Note that for these {constant-$r$}  orbits $E<1$ for $r_0>4m$, $E=1$ for $r_0=4m$,  and $E>1$ for $r_0\in(3m,4m)$. 
Indeed $E\to \infty$ as $r_0\to(3m)^+$. 
\item[(ii)] 
If $\C=0$, then automatically $L=0$, and there is no consistent solution.
\item[(iii)]
If $r_0<2m$ then $X(r_0)$ is a sum of positive and non-negative terms, so there is no consistent solution.
\item[(iv)]
If $r_0=2m$, then from $X(r_0)=0$ we have $E = JL/(4 m^2)$, but then from $X'(r_0)=0$ we have $\C=- 4 m^2 <0$, and so there is no consistent solution.
\item[(v)]
If $r_0= \sqrt[3]{2JL/E}\neq0$ and $r_0> 2m$ then $X(r_0)=0$ implies $1+\C/r_0^2=0$, so that $\C=-r_0^2<0$ and there is no consistent solution.
\item[(vi)]
The generic case is $JL\neq0$, $\C>0$ and $2m < r_0 \neq \sqrt[3]{2JL/E}$.
\end{enumerate}

%\clearpage
Now consider the general case: \\
Choose the three independent variables to be $m$, $E$, 
and $r_0$. Let us solve for $\C(m,E,r_0)$ and $JL(m,E,r_0)$.
First take linear combinations of (\ref{E:X(r)tn}) and (\ref{E:dX(r)tn}) to obtain:
\begin{equation}
3 (E^2-1) r_0^3 +5m r_0^2 - \C (2r_0 -3m) -6 E J L=0;
\end{equation}
\begin{equation}
3 (E^2-1) r_0^6 +4m r_0^5 - \C r_0^4 -12 J^2 L^2=0.
\end{equation}
Solve the first of these equations  for $\C$ to find
\begin{equation}
\C(JL,m,E,r_0) = {3(E^2-1)r_0^3 + 5m r_0^2 - 6 EJL
\over 2r_0-3m}
\end{equation}
Inserting this back into the second equation and solving for $JL$ one finds
\begin{equation}
JL(m,E,r_0) = {\left(Er_0^2 \pm (r_0-2m) \sqrt{r_0\{[9E^2-8]r_0 +12m\}} \right)r_0^2\over 4 (2r_0-3m)} .
\end{equation}

\clearpage
Since $JL$ must be real one in turn deduces $E^2 > {8\over 9} -{4m\over 3 r_0}> {2\over 9}$. However, the \emph{sign} of $JL$ is not constrained; so both roots ($\pm$) are valid. 
%\enlargethispage{30pt}

Inserting this back into $\C(JL,m,E,r_0)$ we see
\begin{eqnarray}
\C(m;E,r_0) &=& { (9E^2-12)r_0^4 -2(9E^2-19)m r_0^3 - 30 m^2 r_0^2 \over 2 (2r_0-3m)^2} 
\nonumber
\\
&&\pm {3Er_0^2(r_0-2m)\sqrt{r_0\{[9E^2-8]r_0 +12m\}} \over 2 (2r_0-3m)^2}.
\end{eqnarray}
The reality conditions for $\C(m;E,r_0)$ are the same as they were for $JL(m;E,r_0)$, that is, $E^2 > {8\over 9} -{4m\over 3 r_0}> {2\over 9}$. To enforce positivity of $\C(m;E,r_0)$, both roots ($\pm$) are acceptable when $E^2 \leq {(3r_0-5m)^2\over9r_0(r_0-2m)}$, but only the + root is acceptable outside this range. 

Then to determine stability one must determine the \emph{sign} of:
\begin{eqnarray}
X''(m; E,r_0) &=& {18E^2(3 r_0^2-10mr_0+9m^2)\over r_0^2(2r_0-3m)^2} 
-{2(12 r_0^2-37mr_0+30m^2)\over r_0^3(2r_0-3m)} 
\nonumber\\&&
\mp {6E(r_0-2m)\sqrt{r_0\{[9E^2-8]r_0 +12m\}} \over r_0^2 (2r_0-3m)^2}.
\end{eqnarray}
That is: 
\begin{eqnarray}
\sign\left(X''(m; E,r_0)\right) &=& \sign\Big\{ {18E^2(3 r_0^2-10mr_0+9m^2)r_0} 
\nonumber\\
&&\qquad
-{2(12 r_0^2-37mr_0+30m^2)(2r_0-3m)} 
\nonumber\\
&&\qquad
\mp {6Er_0(r_0-2m)\sqrt{r_0\{[9E^2-8]r_0 +12m\}}}\Big\}.
\qquad
\end{eqnarray}
In short, there will be \emph{many} {constant-$r$}  orbits, but determining the stability of these {constant-$r$}  orbits as functions of the independent parameters $(m,E,r_0)$ will be extremely tedious.

\vspace{-15pt}
%----------------------------------------------------------
\section{General angular motion for {constant-$r$}  orbits}
%----------------------------------------------------------
\vspace{-10pt}

Now that we have investigated acceptable values of the parameters $\{\C,JL,E,m\}$ and the radius $r_0$ for {constant-$r$}  orbits, we note that  two of the the four constants of the motion reduce to
\begin{equation}\label{Geo_eqn_2b}
\epsilon=  -\left(1-\frac{2m}{r_0} \right) \left(\frac{\d t}{\d\lambda}\right)^2 +{\mathcal{C}\over r_0^2} \ ;
\end{equation}
\begin{equation}\label{E_2b}
E=\left(1-\frac{2m}{r_0}\right)\frac{\d t}{\d\lambda}  + \frac{2JL}{r^3_0} \ .
\end{equation}

\clearpage
Thence for the {constant-$r$}  geodesic trajectories \enlargethispage{30pt} 
\begin{equation}
\frac{\d r}{\d\lambda}= 0 = \frac{\d^2 r}{\d\lambda^2};
\label{E:r(lambda)b}
\end{equation}
\begin{equation}
\frac{\d t}{\d\lambda}= \frac{E-2JL/r_0^3}{(1-2m/r_0)} \ ;
\label{E:tb}
\end{equation}
\begin{equation}
\frac{\d\theta}{\d\lambda}= S_{\theta}\frac{\sqrt{\mathcal{C}-L^2/\sin^2\theta}}{r_0^2} \ ;
\label{E:theta-b}
\end{equation}
\begin{equation}
\frac{\d\phi}{\d\lambda}= \frac{L}{r_0^2\sin^2\theta} + {2J\over r_0^3} \; \frac{E-2JL/r_0^3}{(1-2m/r_0)} \ . \label{E:phi_b}
\end{equation}
We immediately see that $t$ is an affine parameter, that the declination $\theta(\lambda)$  evolves independently of the azimuth $\phi(\lambda)$,  
and that the azimuthal motion depends on a constant drift and a fluctuating term driven by the declination. 
Note that the angular motion is qualitatively unaffected by the difference between timelike and null. 

%----------------------------------------------------------
\section{Declination for {constant-$r$}  orbits ($L\neq 0$)}
%----------------------------------------------------------

Consider the ODE controlling the evolution of the declination $\theta(\lambda)$. 

%\clearpage
%----------------------------------------------------------
\subsection{Forbidden declination range}
%----------------------------------------------------------

The form of the Carter constant, equation \eqref{C_2}, gives a range of forbidden declination angles for any given, non-zero values of $\mathcal{C}$ and $L$. 
We require that $\d\theta/\d\lambda$ be real, and from equation \eqref{C_2} this implies the following requirement:
\begin{equation}
\left(r^2\frac{\d\theta}{\d\lambda}\right)^2=\mathcal{C}-\frac{L^2}{\sin^2\theta}\geq 0
\quad\Longrightarrow\quad
\sin^2 \theta \geq {L^2\over \C} \ .
\end{equation}
Then provided $\C \geq L^2$, which is automatic in view of \eqref{C_2}, we can define a critical angle $\theta_* \in [0,\pi/2]$ by setting 
\begin{equation}
\theta_* = \sin^{-1}(|L|/\sqrt{\C}) \ .
\end{equation}
 Then the allowed range for $\theta$ is the equatorial band:
\begin{equation}
\label{E:theta_range}
\theta \in \Big[ \theta_*, \pi -\theta_*\Big] \ .
\end{equation}
\begin{itemize}
\item 
For $L^2=\C$ we have $\theta=\pi/2$; the motion is restricted to the equatorial plane.
\item
For $L=0$ with $\C>0$ the range of $\theta$ is \emph{a priori} unconstrained; $\theta\in[0,\pi]$.
\item
For $L=0$ with $\C=0$ the declination is fixed $\theta(\lambda)=\theta_{0}$, and the motion is restricted to a constant declination conical surface.
\end{itemize}

\clearpage
%----------------------------------------------------------
\subsection{Evolution of the declination}
%----------------------------------------------------------

As regards the declination angle $\theta$, 
from equation \eqref{E:theta-b}, we find
\begin{eqnarray}
\frac{\d\cos\theta}{\d\lambda} 
&=& -S_{\theta} {\sqrt{\mathcal{C}\sin^2 \theta-L^2}\over r_0^2} 
\nonumber\\
&=& -S_{\theta}\frac{\sqrt{\mathcal{C}}}{r_0^2}\sqrt{\sin^2\theta-\sin^2\theta_*}\;
\nonumber\\
&=& -S_{\theta} \frac{\sqrt{\mathcal{C}}}{r_0^2}\sqrt{\cos^2\theta_*-\cos^2\theta} \ ,
\end{eqnarray}
implying
\begin{equation}
{\d\cos\theta \over  \sqrt{\cos^2\theta_*-\cos^2\theta}}  = -S_{\theta}  \frac{\sqrt{\mathcal{C}}}{r_0^2}\,\d\lambda
 \ .
\end{equation}

\begin{figure}[!htbp]
\begin{center}
\includegraphics[scale=0.5]{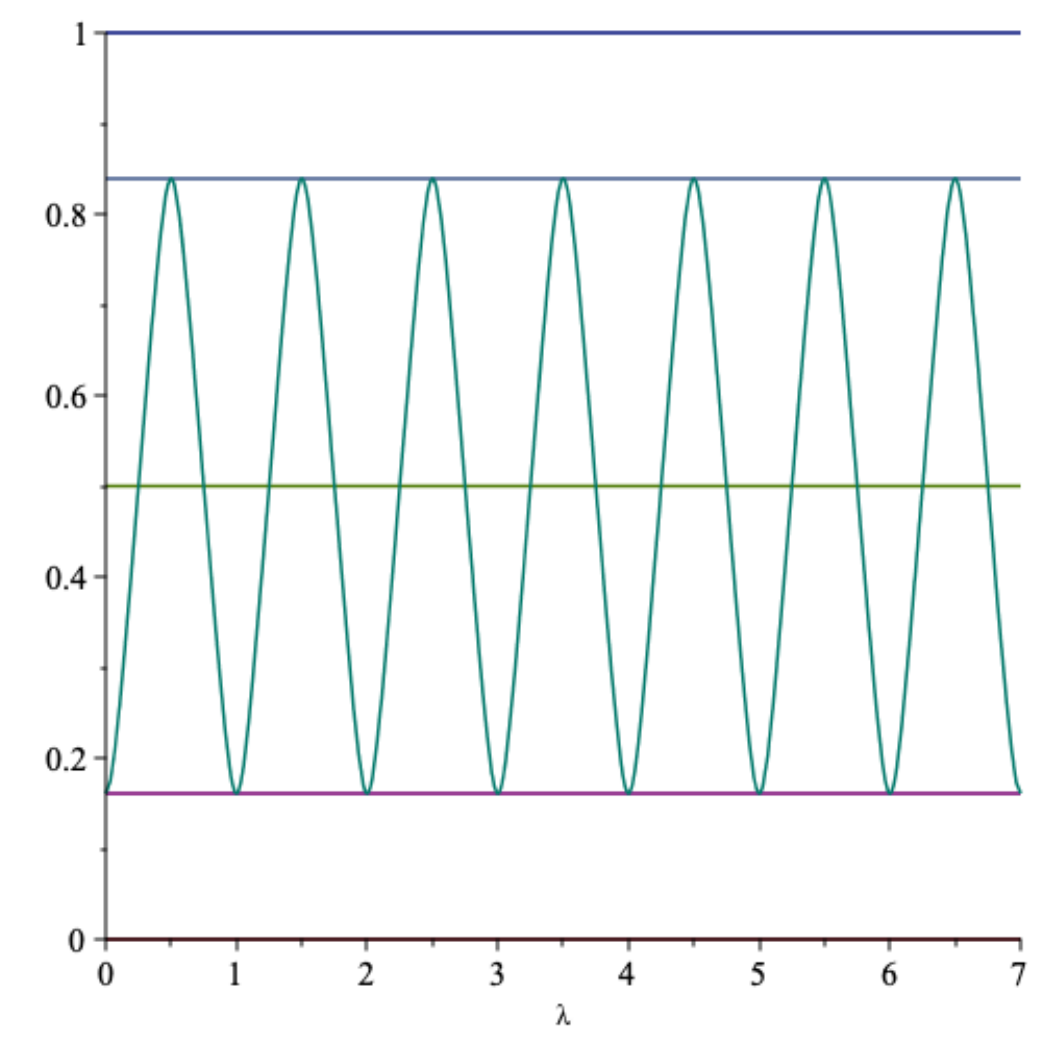}
\caption{Qualitative behaviour of the declination $\theta(\lambda)/\pi$ as it oscillates back and forth between $\theta_*/\pi$ and $(\pi-\theta_*)/\pi$.}
\label{F:declination}
\end{center}
\end{figure}

%\clearpage
From this we see
\begin{equation}
{\d\cos^{-1}\left(\frac{\cos\theta}{\cos\theta_*}\right)} = S_{\theta}\frac{\sqrt{\C}}{r_0^2}\d\lambda \ ,
\end{equation}
that is
\begin{equation}
\label{E:declination}
 {\cos^{-1}\left(\frac{\cos\theta}{\cos\theta_*} \right)} =  {\cos^{-1}\left(\frac{\cos\theta_0}{\cos\theta_*} \right)} + S_{\theta}\frac{\sqrt{\C}}{r_0^2} (\lambda-\lambda_0) \ .
\end{equation}
Without loss of generality we may allow the geodesic to reach the critical angle $\theta_*$ at some affine parameter $\lambda_*$, and then use that as our new initial data. This effectively sets $\theta_0=\theta_*$, and gives us the following simple result:
\begin{equation}
\label{E:declination2}
    \cos^{-1}\left(\frac{\cos\theta}{\cos\theta_*}\right) = 
    S_{\theta}\frac{\sqrt{\C}}{r_0^2}(\lambda-\lambda_*)  \ .
\end{equation}
Thence, using the fact that cosine is an even function of its argument:
\begin{eqnarray}
\label{E:declination3}
    \cos\theta &=& \cos\theta_*\;\cos\left(S_{\theta}\frac{\sqrt{\C}}{r_0^2}(\lambda-\lambda_*) \right) 
    = \cos\theta_*\;\cos\left(\frac{\sqrt{\C}}{r_0^2}(\lambda-\lambda_*) \right) \ . \nonumber
\end{eqnarray}
{For a qualitative plot of the declination angle as a function of affine parameter see figure~\ref{F:declination}. }
Note the motion is periodic, with period 
\begin{equation}
\Delta \lambda = {2\pi r_0^2\over\sqrt{\C}}.
\end{equation}
In terms of the Killing time coordinate the period is
\begin{equation}
T_\theta = \frac{2\pi(E-2JL/r_0^3)}{\sqrt{\C}(1-2m/r_0)}.
\end{equation}

%==============================
\section{Azimuth for {constant-$r$}  orbits ($L\neq0$)}
\label{SS:azimuth}
%==============================

Now consider the ODE for the evolution of the azimuthal angle $\phi(\lambda)$. 
We have
\begin{equation}
\label{E:8.8}
\frac{d\phi}{d\lambda}= \frac{2J}{r_0^3}\; \left({E-2JL/r_0^3\over1-2m/r_0}\right) + { L\over r_0^2 \sin^2\theta} \ ,
\end{equation}
and
\begin{eqnarray}
\label{E:declination3_b}
    \cos\theta &=& 
    \cos\theta_*\;\cos\left(\frac{\sqrt{\C}}{r_0^2}(\lambda-\lambda_*) \right) \ . \nonumber
\end{eqnarray}
Thence
\begin{equation}
\phi(\lambda) = \phi_* + \frac{2J}{r_0^3}\; \left({E-2JL/r_0^3\over1-2m/r_0}\right) (\lambda-\lambda_*) 
+ {L\over r_0^2} \int_{\lambda_*}^\lambda 
{\d\bar\lambda\over 1-\cos^2{\theta(\bar\lambda)}}.
\end{equation}
The only tricky item here is evaluation of the integral
\begin{equation}
\int_{\lambda_*}^\lambda 
{\d\bar\lambda\over 1-\cos^2{\theta(\bar\lambda)}}
=
\int_{\lambda_*}^\lambda 
{\d\bar\lambda\over 1-\cos^2(\theta_*) \cos^2\left( (\sqrt{\C}/ r_0^2) (\bar\lambda-\lambda_*)\right)}.
\end{equation}

\clearpage
But it is easy to check that {formally}
\begin{equation}
 \int {d\lambda \over 1- (A \cos( B+ F\lambda))^2} 
=  {1\over F \sqrt{1-A^2}} \;\arctan\left(\tan(B+F\lambda)\over\sqrt{1-A^2}\right) \ .
\end{equation}

Note that the LHS above is monotone increasing, while the RHS naively exhibits discontinuities whenever the tangent passes through infinity. Thence the correct statement is to observe that
\begin{eqnarray}
 \int {d\lambda \over 1- (A \cos( B+ F\lambda))^2} 
&=&  {1\over F \sqrt{1-A^2}} \;\Bigg\{ \arctan\left(\tan(B+F\lambda)\over\sqrt{1-A^2}\right) \nonumber\\
&& \qquad\qquad\qquad  + \pi \; \floor\left[{ B+ F\lambda+\pi/2\over\pi} \right] \Bigg\}.
\end{eqnarray}
Here $\floor[...]$ denotes the integer part (floor function) and the discontinuity in $\floor[...]$ exactly cancels the discontinuity due to the $\arctan(...)$. In terms of the fractional part function $\Frac[...]$ one has $x = \floor[x] + \Frac[x]$ so one could equally well write 
\begin{eqnarray}
 \int {d\lambda \over 1- (A \cos( B+ F\lambda))^2} 
&=&  {1\over F \sqrt{1-A^2}} \;\Bigg\{ \arctan\left(\tan(B+F\lambda)\over\sqrt{1-A^2}\right) \nonumber\\
&&  - \pi \; \Frac\left[{ B+ F\lambda+\pi/2\over\pi} \right]  + B+ F\lambda+\pi/2 \Bigg\}.
\end{eqnarray}
\enlargethispage{20pt}

Thence
\begin{eqnarray}
\int_{\lambda_*}^\lambda 
{\d\bar\lambda\over \sin^2(\theta(\bar\lambda))}
&=&
{r_0^2\over \sqrt{\C} \sin\theta_*} {\Bigg\{ }
\arctan\left(\tan(({\sqrt{\C}}/ r_0^2) (\lambda-\lambda_*))\over\sin\theta_*\right)
\\
&& 
 - \pi \; \Frac\left[{(\sqrt{\C}/ r_0^2) (\lambda-\lambda_*)+\pi/2\over\pi} \right]  + (\sqrt{\C}/ r_0^2) (\lambda-\lambda_*)+\pi/2 \Bigg\}.\;\;
 \nonumber
\end{eqnarray}
Finally, using $\C = L^2/\sin^2\theta{_*}$, we have
\begin{eqnarray}
\phi &=& \phi_*  +  {\Bigg\{} \frac{2J}{r_0^3}\; {E-2JL/r_0^3\over1-2m/r_0}{ +{\sqrt{\C}\over r_0^2}  \Bigg\} } (\lambda-\lambda_*)
 + \arctan\left({1\over\sin\theta_*}
 \tan\left( \sqrt{\C} \;{\lambda-\lambda_*\over r_0^2}\right)\right)
\nonumber \\
 &&
 \qquad\qquad\qquad
  - \pi \; \Frac\left[{(\sqrt{\C}/ r_0^2) (\lambda-\lambda_*)+\pi/2\over\pi} \right]  +{\pi\over 2}.
 \end{eqnarray}
 So the azimuthal motion is a constant drift (growing linearly in the affine parameter) with a superimposed oscillation. 
 
\begin{figure}[!htbp]
\begin{center}
\includegraphics[scale=0.50]{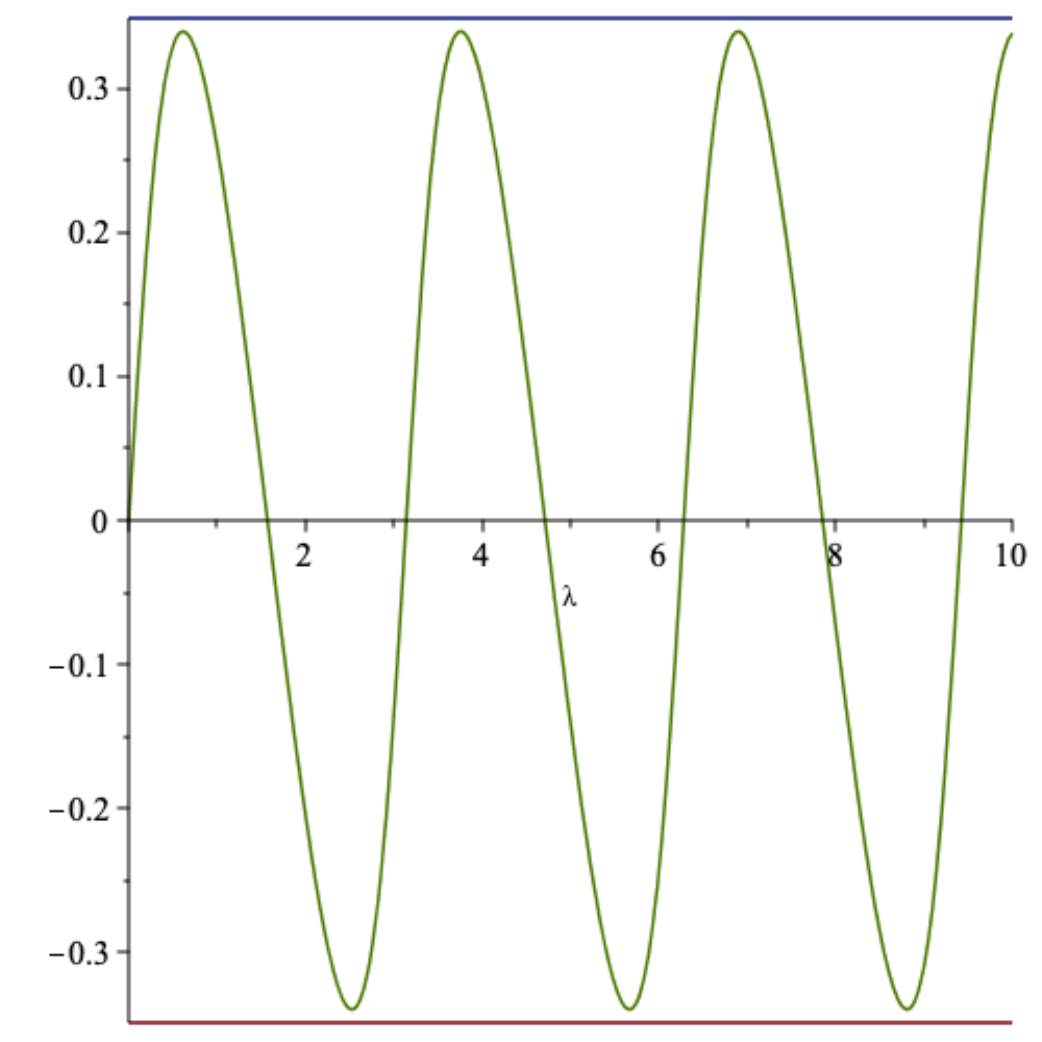}
\caption{Qualitative behaviour of the oscillatory part $\phi_{oscillation}(\lambda)$ of the azimuthal evolution as a function of the affine parameter. }
\label{F:azimuth}
\end{center}
\end{figure}

\begin{figure}[!htbp]
\begin{center}
\includegraphics[scale=0.50]{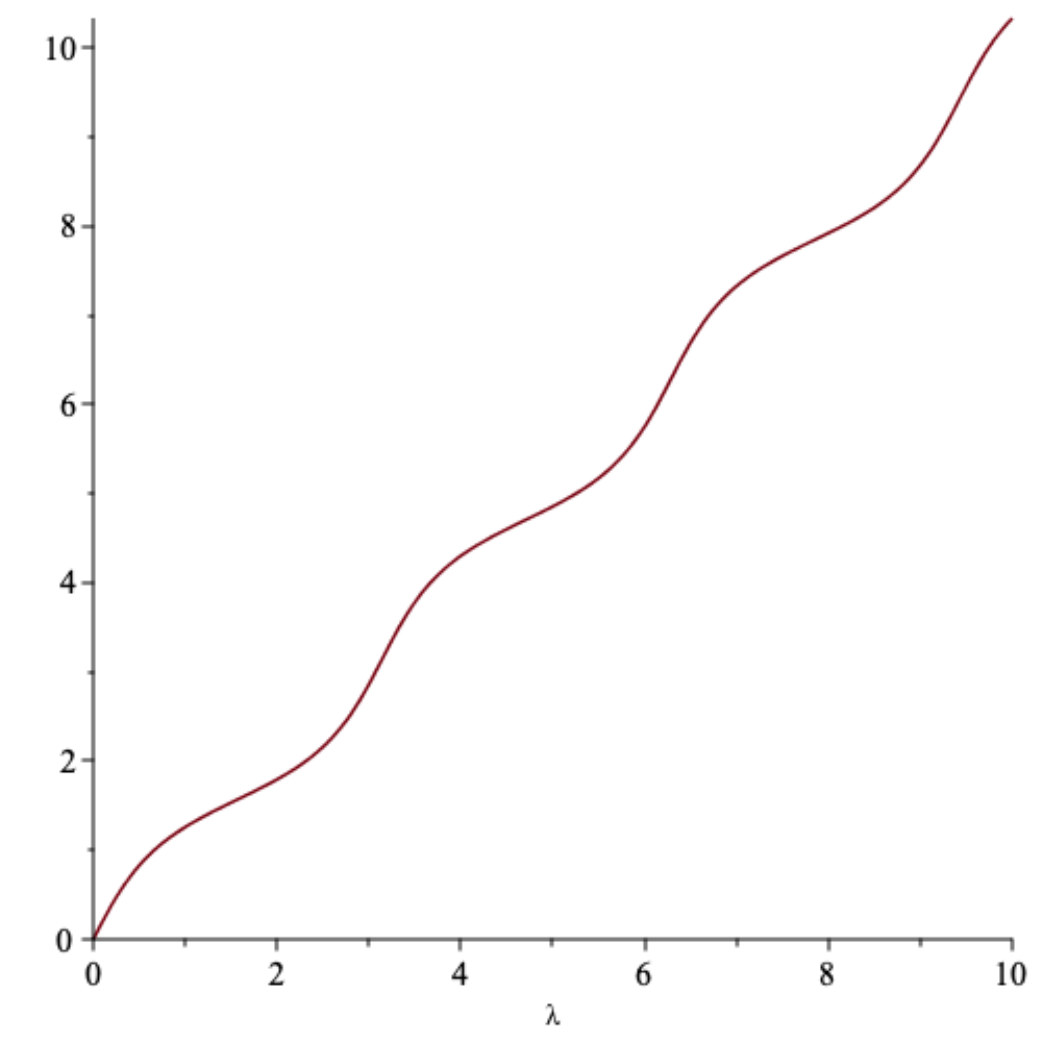}
\caption{Qualitative behaviour of  $\phi(\lambda)$, the total  azimuthal phase evolution (drift plus oscillation) as a function of the affine parameter. }
\label{F:phase}
\end{center}
\end{figure}

 \clearpage
Specifically the oscillatory term is
 \begin{eqnarray}
\phi_{oscillation}(\lambda) &=&{
\arctan\left({1\over\sin\theta_*}
 \tan\left( \sqrt{\C} \;{\lambda-\lambda_*\over r_0^2}\right)\right)}
 \nonumber\\
 &&
  - \pi \; \Frac\left[{(\sqrt{\C}/ r_0^2) (\lambda-\lambda_*)+\pi/2\over\pi} \right]  +{\pi\over 2}.
\end{eqnarray}
Note the sensible limit for equatorial motion as $\sin\theta_*\to 1$.
See figure~\ref{F:azimuth} for a qualitative plot of the oscillating term, and figure~\ref{F:phase} for a qualitative plot of the total phase (drift plus oscillation).

\clearpage
The oscillatory contribution to the azimuthal evolution has the same period as the evolution in declination 
 \begin{equation}
\Delta \lambda_{oscillation} = {2\pi r_0^2\over\sqrt{\C}},
\end{equation}
but the drift component has periodicity
\begin{equation}
\Delta \lambda_{drift} = 
2\pi {\Bigg\{} \frac{2J}{r_0^3}\; {E-2JL/r_0^3\over1-2m/r_0} { +{\C\over r_0^2}  \Bigg\}^{-1} }
\end{equation}
so that
\begin{equation}
(T_\phi)_{drift} = 
2\pi \; {E-2JL/r_0^3\over1-2m/r_0} \; {\Bigg\{} \frac{2J}{r_0^3}\; {E-2JL/r_0^3\over1-2m/r_0}{ +{\C\over r_0^2}  \Bigg\}^{-1} }
\end{equation}
This drift in azimuth periodicity is typically incommensurate with the periodicity in declination, so the geodesics are surface filling and will eventually cover the entire equatorial band $\theta\in[\theta_*,\pi-\theta_*]$. ({See figure~\ref{F:band} for a qualitative description.})

%\clearpage
%----------------------------------------------------------------------------
\section{Angular motion for $L=0$ ({Constant-$r$}  ZAMOS)}
%----------------------------------------------------------------------------

If we now consider the special case of {constant-$r$}  orbits where $L=0$, then $\sin\theta_*\to 0$,\\ so we need to be careful. The equations of motion reduce even further to:
\begin{equation}
\left(\frac{d\phi}{d\lambda}\right) = \frac{2J}{r_0^3}\frac{E}{1-2m/r_0} \ ;
\qquad
\left(\frac{d\theta}{d\lambda}\right) =  \pm{\sqrt{\mathcal{C}}\over r_0^2} \ .
\end{equation}
So in this special case we find
\begin{equation}
\phi=\phi_0 + \frac{2J}{r_0^3}\frac{E}{1-2m/r_0}(\lambda-\lambda_0) \ ;
\qquad
\theta=\theta_0\pm  {\sqrt{\mathcal{C}}\over r_0^2} (\lambda-\lambda_0) \ .
\end{equation}
Now $\phi$ is defined only modulo $2\pi$, but $\theta$ is naively in $[0,\pi]$. However if we formally drive it outside this range we just need to reset $\phi$ by $\pi$.
That is, we can identify the points $(\theta+\pi,\phi) \equiv (\pi-\theta, \phi+\pi)$.

In view of the fact that for {constant-$r$}  orbits with $L=0$ the quantity
\begin{equation}
{dt\over d\lambda} = {E\over1-2m/r_0} 
\end{equation}
is a constant, we can also rewrite angular dependence as
\begin{equation}
\phi=\phi_0 + \frac{2J}{r_0^3}(t-t_0) \ ;
\qquad
\theta=\theta_0\pm  {\sqrt{\mathcal{C}}(1-2m/r_0)\over E r_0^2}\; (t-t_0) \ .
\end{equation}%\enlargethispage{30pt}

Note the periodicities in azimuth and declination are
%\enlargethispage{20pt}
\begin{equation}
T_\phi = {\pi r_0^3\over J},
\qquad \hbox{and} \qquad
T_\theta= {\pi E r_0^2\over \sqrt{\mathcal{C}}(1-2m/r_0)}.
\end{equation}
These are typically incommensurate, so these ZAMO curves are surface filling and will eventually cover the entire angular 2-sphere. ({The equatorial band in figure~\ref{F:band} will expand to include both poles.})

%----------------------------------------------------------------
\section{Limit as $J\to 0$}
%----------------------------------------------------------------

Physically the limit $J\to 0$ corresponds to switching off the angular momentum of the central object generating the gravitational field, so that the spacetime becomes Schwarzschild in Painlev\'e--Gullstrand coordinates; so for {constant-$r$}  orbits we must recover the unstable photon sphere at $r=3m$ and the ISCO at $r=6m$.  If not, something is very wrong.

For $J\to0$ the quantity $X(r)$ simplifies to 
\begin{equation} \label{X2}
X(r) \to E^2-\left(1-\frac{2m}{r}\right)\left(-\epsilon + \frac{\mathcal{C}}{r^2}\right) \ .
\end{equation}

%-------------------------------------------------
\subsection{Photon spheres}
%--------------------------------------
For massless particles $\epsilon\to 0$, and without loss of generality we can set $E\to 1$. 
This implies
\begin{equation} \label{X0}
X(r) \to 1-\left(1-\frac{2m}{r}\right) \frac{\mathcal{C}}{r^2};
\end{equation}
\begin{equation} \label{X1b}
X'(r) \to {2\C (r-3m)\over r^4};
\end{equation}
and
\begin{equation} \label{X2b}
X''(r) \to -\,{6\C (r-4m)\over r^5}.
\end{equation}

%\clearpage
There is a unique photon sphere at $r_0=3m$. Then $X(r_0) = 0 = 1 -\C/(3 r_0^2)$, that is $\C=3 r_0^2$. We then see that $X''(3m) = 2\C/(243 m^4) = 2/(81 m^2)> 0$, these photon orbits are unstable. This is exactly as it should be.

%-------------------------------------------------
\subsection{Massive particle spheres}
%-------------------------------------------------
For massive particles $\epsilon\to -1$, that implies
\begin{equation} \label{XX0}
X(r) \to E ^2-\left(1-\frac{2m}{r}\right) \left( 1 + \frac{\mathcal{C}}{r^2}\right);
\end{equation}
\begin{equation} \label{XX1}
X'(r) \to {2\C (r-3m)-2mr^2\over r^4}.
\end{equation}
and 
\begin{equation} \label{XX2}
X''(r) \to \,{-6\C (r-4m) +4mr^2\over r^5}.
\end{equation}\enlargethispage{40pt}
Solve $X'(r_0)=0$ to find $\C(m,r_0)$:
\begin{equation}
\C(m,r_0)= {m r_0^2\over r_0-3m}.
\end{equation}
Since $\C \geq 0$, there will now be many {constant-$r$}  orbits, all the way from $r_0=\infty$ down to $r_0=3m$. 
Use this to evaluate $X''(m,r_0)$:
 \begin{equation} \label{XX2b}
X''(r_0,m) \to -\,{2m(r_0-6m)\over r_0^3(r_0-3m)}.
\end{equation}

%\enlargethispage{20pt}
%\clearpage
Inspecting the \emph{sign} of $X''(m,r_0)$, the {constant-$r$}  orbits are stable for $r_0>6m$, marginal for $r_0=6m$, and unstable for $r_0<6m$.  This is exactly as it should be.

%\clearpage
%=================================
\section{Conclusions}
\label{S:conclusions}
%=================================

We have explored the existence of and properties of the {constant-$r$} (``quasi-circular'') geodesics in the recently introduced Painlev\'e--Gullstand variant of the Lense--Thirring spacetime~\cite{PGLT1,PGLT2,PGLT3}. 
We emphasize that although the underlying spacetime is not spherically symmetric, (only stationary and axisymmetric), so that the Birkhoff theorem does not apply~\cite{Birkhoff, Jebsen, Deser, Ravndal, Skakala}, one nevertheless encounters (partial) spherical shells of {constant-$r$}  geodesics; notably this behaviour is not limited to the (exact) Kerr spacetime, but also persists in the Painlev\'e--Gullstand variant of the Lense--Thirring spacetime. 
{The persistence of existence of these {constant-$r$} (``quasi-circular'') geodesics is intimately related to the persistence of existence of a non-trivial Killing tensor and the  associated Carter constant.} 
Overall, we see that the Painlev\'e--Gullstand variant of the Lense--Thirring spacetime~\cite{PGLT1,PGLT2,PGLT3} exhibits many useful and interesting 
properties, and is well-adapted to direct confrontation with observational astrophysics. 

\clearpage
From a wider perspective, these considerations can be viewed as an element of the study of modified black holes --- alternative black holes to the standard Schwarzschild--Kerr family that are nevertheless carefully formulated so as to pass the most obvious observational tests,  and so provide useful templates for  driving observational astrophysics~\cite{Carballo-Rubio:2018, Carballo-Rubio:2019,  
Carballo-Rubio:2021, Pandora, Simpson:2021a, Simpson:2021b, Visser:2009a, Visser:2009b, bb-Kerr,bb-KN,DeLorenzo:2014,Hayward:2005, Bronnikov:2000, Bronnikov:2005, Johannsen:2011, Bardeen:1968}. 

%\enlargethispage{30pt}
%\clearpage
%===================================
\section*{Acknowledgements}
%===================================

JB was supported by a MSc scholarship funded by the Marsden Fund, 
via a grant administered by the Royal Society of New Zealand.
\\
TB was supported by a Victoria University of Wellington MSc scholarship, 
and was also indirectly supported by the Marsden Fund, 
via a grant administered by the Royal Society of New Zealand.
\\
AS was supported by a Victoria University of Wellington PhD Doctoral Scholarship,
and was also indirectly supported by the Marsden fund, 
via a grant administered by the Royal Society of New Zealand.
\\
MV was directly supported by the Marsden Fund, \emph{via} a grant administered by the Royal Society of New Zealand.

%\clearpage
\bigskip
%===================================
\hrule\hrule\hrule
%=================================== 

%=================================

\begin{thebibliography}{99}
%===================================
\newcommand{\arXiv}[1]{arXiv:\href{https://arxiv.org/abs/#1}{\color{blue}#1}}
%This allows using \arXiv{2006.07125} or \arXiv{gr-qc/0009013} for nice links.
%==================================
%===================================

\bibitem{Kerr}
Roy Kerr, \\
``Gravitational field of a spinning mass as an example of algebraically special metrics'',
 Physical Review Letters {\bf  11} 237-238 (1963).

\bibitem{Kerr-Texas}
Roy Kerr,\\
``Gravitational collapse and rotation'',\\
published in: 
{\sl Quasi-stellar sources and gravitational collapse:
Including the proceedings of the First Texas Symposium on Relativistic
Astrophysics}, edited by Ivor Robinson, Alfred Schild, and E.L. Sch\"ucking
(University of Chicago Press, Chicago, 1965), pages 99--102.\\
The conference was held in Austin, Texas, on 16--18 December 1963.

\bibitem{kerr-newman}
E. Newman, E. Couch, K. Chinnapared, A. Exton, A. Prakash and R. Torrence,
``Metric of a Rotating, Charged Mass'',
J. Math. Phys. \textbf{6} (1965) 918.

\bibitem{kerr-intro}
M.~Visser,
``The Kerr spacetime: A brief introduction'',
[\arXiv{0706.0622} [gr-qc]].
Published in \cite{kerr-book}.
%94 citations counted in INSPIRE as of 13 Jun 2020


%\clearpage
\bibitem{kerr-book}
D.~L.~Wiltshire, M.~Visser and S.~M.~Scott (editors),\\
\emph{The Kerr spacetime: Rotating black holes in general relativity},\\
(Cambridge University Press, Cambridge, 2009).
%31 citations counted in INSPIRE as of 13 Jun 2020


%\clearpage
\bibitem{kerr-book-2}
Barrett O'Neill,
\emph{The geometry of Kerr black holes},\\
(Peters, Wellesley, 1995). Reprinted (Dover, Mineloa, 2014).

%================================================

\bibitem{MTW}
Charles Misner, Kip Thorne, and John Archibald Wheeler,  {\sl Gravitation}, \\
(Freeman, San Francisco, 1973).

\bibitem{Wald}
Robert Wald,
\emph{General relativity},\\
(University of  Chicago Press, Chicago, 1984).

\bibitem{Weinberg}
Steven Weinberg,\\
\emph{Gravitation and Cosmology: Principles and Applications of the General Theory of Relativity},
(Wiley,  Hoboken, 1972).

\bibitem{Adler-Bazin-Schiffer}
Ronald J. Adler, Maurice Bazin, and Menahem Schiffer, \\
 {\sl Introduction to General Relativity}, Second edition, \\
 (McGraw--Hill, New York, 1975).\\
 {}[It is important to acquire the 1975 second edition, the 1965 first edition does not contain any discussion of the Kerr spacetime.]
 
\bibitem{Hobson}
M.~P.~Hobson, G.~P.~Estathiou, and A~N.~Lasenby,\\
\emph{General relativity: An introduction for physicists},\\
(Cambridge University Press, Cambridge, 2006).

\bibitem{D'Inverno}
Ray D'Inverno, {\sl Introducing Einstein's Relativity}, (Oxford University Press, 1992).

\bibitem{Hartle}
James Hartle, {\sl Gravity: An introduction to Einstein's general relativity},\\
(Addison Wesley, San Francisco, 2003).

%\clearpage
\bibitem{Carroll}
Sean Carroll, {\sl  An introduction to general relativity: Spacetime and Geometry},
(Addison Wesley, San Francisco, 2004).


%================================================
\bibitem{Berti:2015}
E.~Berti, E.~Barausse, V.~Cardoso, L.~Gualtieri, P.~Pani, U.~Sperhake, L.~C.~Stein, N.~Wex, K.~Yagi and T.~Baker, \textit{et al.}
``Testing General Relativity with Present and Future Astrophysical Observations'',
Class. Quant. Grav. \textbf{32} (2015), 243001
\doi{10.1088/0264-9381/32/24/243001}
[\arXiv{1501.07274} [gr-qc]].
%862 citations counted in INSPIRE as of 10 Feb 2022

\bibitem{Yunes:2016}
N.~Yunes, K.~Yagi and F.~Pretorius,
``Theoretical Physics Implications of the Binary Black-Hole Mergers GW150914 and GW151226'',
Phys. Rev. D \textbf{94} (2016) no.8, 084002
\doi{10.1103/PhysRevD.94.084002}
[\arXiv{1603.08955} [gr-qc]].
%483 citations counted in INSPIRE as of 10 Feb 2022

\bibitem{Cardoso:2019}
V.~Cardoso and P.~Pani,
``Testing the nature of dark compact objects: a status report'',
Living Rev. Rel. \textbf{22} (2019) no.1, 4
\doi{10.1007/s41114-019-0020-4}
[\arXiv{1904.05363} [gr-qc]].
%307 citations counted in INSPIRE as of 10 Feb 2022

\bibitem{Barack:2006}
L.~Barack and C.~Cutler,
``Using LISA EMRI sources to test off-Kerr deviations in the geometry of massive black holes'',
Phys. Rev. D \textbf{75} (2007), 042003
\doi{10.1103/PhysRevD.75.042003}
[\arXiv{gr-qc/0612029} [gr-qc]].
%185 citations counted in INSPIRE as of 10 Feb 2022

\bibitem{Bambi:2019}
C.~Bambi, K.~Freese, S.~Vagnozzi and L.~Visinelli,
``Testing the rotational nature of the supermassive object M87* from the circularity and size of its first image'',
Phys. Rev. D \textbf{100} (2019) no.4, 044057
\doi{10.1103/PhysRevD.100.044057}
[\arXiv{1904.12983} [gr-qc]].
%151 citations counted in INSPIRE as of 10 Feb 2022

\bibitem{Barack:2018}
L.~Barack and A.~Pound,
``Self-force and radiation reaction in general relativity'',
Rept. Prog. Phys. \textbf{82} (2019) no.1, 016904
\doi{10.1088/1361-6633/aae552}
[\arXiv{1805.10385} [gr-qc]].
%137 citations counted in INSPIRE as of 10 Feb 2022

\bibitem{LISA:2020}
E.~Barausse, E.~Berti, T.~Hertog, S.~A.~Hughes, P.~Jetzer, P.~Pani, T.~P.~Sotiriou, N.~Tamanini, H.~Witek and K.~Yagi, \textit{et al.}
``Prospects for Fundamental Physics with LISA'',
Gen. Rel. Grav. \textbf{52} (2020) no.8, 81
\doi{10.1007/s10714-020-02691-1}
[\arXiv{2001.09793} [gr-qc]].
%123 citations counted in INSPIRE as of 10 Feb 2022

%================================================



%================================================


\bibitem{Birkhoff}
Garret Birkhoff,  % G. D. ,
 \emph{Relativity and Modern Physics}, 
(Harvard University Press, Cambridge, 1923).% LCCN 23008297. 

\bibitem{Jebsen}
J\o{}rg Tofte Jebsen, ``\"Uber die allgemeinen kugelsymmetrischen 
L\"osungen der Einsteinschen Gravitationsgleichungen im Vakuum'', 
Ark. Mat. Ast. Fys. (Stockholm) {\bf 15} (1921) nr.18. 

\bibitem{Deser}
  Stanley Deser and Joel Franklin,
  ``Schwarzschild and Birkhoff \emph{a la} Weyl'',\\
  Am.\ J.\ Phys.\  {\bf 73} (2005) 261
  [\arXiv{gr-qc/0408067} [gr-qc]].
  %%CITATION = AJPIA,73,261;%%
  
 \bibitem{Ravndal} 
   Nils Voje Johansen, Finn Ravndal, 
   ``On the discovery of Birkhoff's theorem'', 
   Gen.Rel.Grav. 38 (2006) 537-540  
   [\arXiv{physics/0508163} [physics.hist-ph]].
 
\bibitem{Skakala}
J.~Skakala and M.~Visser,\\
``Birkhoff-like theorem for rotating stars in (2+1) dimensions'',\\{}
[\arXiv{0903.2128} [gr-qc]].
%4 citations counted in INSPIRE as of 24 Jun 2020

%\clearpage
%================================================
\bibitem{Edery:2006}
A.~Edery and J.~Godin,
``Second order Kerr deflection'',\\
Gen. Rel. Grav. \textbf{38} (2006), 1715-1722
\doi{10.1007/s10714-006-0347-5}
%15 citations counted in INSPIRE as of 03 Mar 2022

%================================================
%\clearpage
\bibitem{Hod:2011}
S.~Hod,
``The fastest way to circle a black hole'',
Phys. Rev. D \textbf{84} (2011), 104024
\doi{10.1103/PhysRevD.84.104024}
[\arXiv{1201.0068} [gr-qc]].
%42 citations counted in INSPIRE as of 03 Dec 2021

\bibitem{Warburton:2013}
N.~Warburton, L.~Barack and N.~Sago,
``Isofrequency pairing of geodesic orbits in Kerr geometry'',
Phys. Rev. D \textbf{87} (2013) no.8, 084012
\doi{10.1103/PhysRevD.87.084012}
[\arXiv{1301.3918} [gr-qc]].
%33 citations counted in INSPIRE as of 03 Dec 2021

\bibitem{Hod:2012}
S.~Hod,
``Spherical null geodesics of rotating Kerr black holes'',
Phys. Lett. B \textbf{718} (2013), 1552-1556
\doi{10.1016/j.physletb.2012.12.047}
[\arXiv{1210.2486} [gr-qc]].
%32 citations counted in INSPIRE as of 03 Dec 2021

\bibitem{Teo:2020}
E.~Teo,
``Spherical orbits around a Kerr black hole'',
Gen. Rel. Grav. \textbf{53} (2021) no.1, 10
\doi{10.1007/s10714-020-02782-z}
[\arXiv{2007.04022} [gr-qc]].
%16 citations counted in INSPIRE as of 03 Dec 2021


\bibitem{Tavlayan:2020}
A.~Tavlayan and B.~Tekin,
``Exact Formulas for Spherical Photon Orbits Around Kerr Black Holes'',
Phys. Rev. D \textbf{102} (2020) no.10, 104036
\doi{10.1103/PhysRevD.102.104036}
[\arXiv{2009.07012} [gr-qc]].
%1 citations counted in INSPIRE as of 04 Dec 2021

%================================================
% photon rings and silhouettes....
%=================================
\bibitem{Broderick:2013}
A.~E.~Broderick, T.~Johannsen, A.~Loeb and D.~Psaltis,
``Testing the No-Hair Theorem with Event Horizon Telescope Observations of Sagittarius A*'',
Astrophys. J. \textbf{784} (2014), 7
\doi{10.1088/0004-637X/784/1/7}
[\arXiv{1311.5564} [astro-ph.HE]].
%159 citations counted in INSPIRE as of 17 Feb 2022

\enlargethispage{20pt}
\bibitem{Johannsen:2016}
T.~Johannsen,
``Sgr A* and General Relativity'',
Class. Quant. Grav. \textbf{33} (2016) no.11, 113001
\doi{10.1088/0264-9381/33/11/113001}
[\arXiv{1512.03818} [astro-ph.GA]].
%104 citations counted in INSPIRE as of 17 Feb 2022

%\clearpage
\bibitem{Broderick:2008}
A.~Broderick and A.~Loeb,
``Imaging the Black Hole Silhouette of M87: Implications for Jet Formation and Black Hole Spin'',
Astrophys. J. \textbf{697} (2009), 1164-1179
\doi{10.1088/0004-637X/697/2/1164}
[\arXiv{0812.0366} [astro-ph]].
%89 citations counted in INSPIRE as of 17 Feb 2022




\bibitem{Gralla:2019}
S.~E.~Gralla, D.~E.~Holz and R.~M.~Wald,
``Black Hole Shadows, Photon Rings, and Lensing Rings'',
Phys. Rev. D \textbf{100} (2019) no.2, 024018
\doi{10.1103/PhysRevD.100.024018}
[\arXiv{1906.00873} [astro-ph.HE]].
%130 citations counted in INSPIRE as of 17 Feb 2022


\bibitem{Pappas:2018}
K.~Glampedakis and G.~Pappas,
``Modification of photon trapping orbits as a diagnostic of non-Kerr spacetimes'',
Phys. Rev. D \textbf{99} (2019) no.12, 124041
\doi{10.1103/PhysRevD.99.124041}
[\arXiv{1806.09333} [gr-qc]].
%27 citations counted in INSPIRE as of 11 Jun 2022



\bibitem{Perlick:2021}
V.~Perlick and O.~Y.~Tsupko,
``Calculating black hole shadows: Review of analytical studies'',
Phys. Rept. \textbf{947} (2022), 1-39
\doi{10.1016/j.physrep.2021.10.004}
[\arXiv{2105.07101} [gr-qc]].
%59 citations counted in INSPIRE as of 11 Jun 2022





% off-axis accretion disks...

\bibitem{Bambi:2017}
C.~Bambi,
``Astrophysical Black Holes: A Compact Pedagogical Review'',
Annalen Phys. \textbf{530} (2018), 1700430
\doi{10.1002/andp.201700430}
[\arXiv{1711.10256} [gr-qc]].
%62 citations counted in INSPIRE as of 17 Feb 2022



\bibitem{Vincent:2020}
F.~H.~Vincent, M.~Wielgus, M.~A.~Abramowicz, E.~Gourgoulhon, J.~P.~Lasota, T.~Paumard and G.~Perrin,
``Geometric modeling of M87* as a Kerr black hole or a non-Kerr compact object'',
Astron. Astrophys. \textbf{646} (2021), A37
\doi{10.1051/0004-6361/202037787}
[\arXiv{2002.09226} [gr-qc]].
%35 citations counted in INSPIRE as of 17 Feb 2022








\bibitem{Chael:2021}
A.~Chael, M.~D.~Johnson and A.~Lupsasca,
``Observing the Inner Shadow of a Black Hole: A Direct View of the Event Horizon'',
Astrophys. J. \textbf{918} (2021) no.1, 6
\doi{10.3847/1538-4357/ac09ee}
[\arXiv{2106.00683} [astro-ph.HE]].
%20 citations counted in INSPIRE as of 17 Feb 2022

\bibitem{Berry:ISCOs}
T.~Berry, A.~Simpson and M.~Visser,
``Photon spheres, ISCOs, and OSCOs: Astrophysical observables for regular black holes with asymptotically Minkowski cores'',
Universe \textbf{7} (2020) no.1, 2
\doi{10.3390/universe7010002}
[\arXiv{2008.13308} [gr-qc]].
%13 citations counted in INSPIRE as of 17 Feb 2022

%\clearpage
\bibitem{Boonserm:2019}
P.~Boonserm, T.~Ngampitipan, A.~Simpson and M.~Visser,
``Innermost and outermost stable circular orbits in the presence of a positive cosmological constant'',
Phys. Rev. D \textbf{101} (2020) no.2, 024050
\doi{10.1103/PhysRevD.101.024050}
[\arXiv{1909.06755} [gr-qc]].
%13 citations counted in INSPIRE as of 17 Feb 2022

%================================================
\bibitem{Lense-Thirring}
Hans Thirring and Josef Lense,  ``\"Uber den Einfluss der Eigenrotation der Zentralk\"orperauf die Bewegung 
der Planeten und Monde nach der Einsteinschen Gravitationstheorie'', Physikalische Zeitschrift, Leipzig Jg. {\bf 19}  (1918), No. 8, p. 156--163.\\ 
%[204--205].\\
English translation by Bahram Mashoon, Friedrich W. Hehl, and Dietmar S. Theiss: ``On the influence of the proper rotations of central bodies on the motions of planets and moons in Einstein's theory of gravity'', General Relativity and Gravitation  {\bf 16} (1984) 727--741.
%[711].

\bibitem{Pfister}
Herbert Pfister, ``On the history of the so-called Lense--Thirring effect'',
\url{http://philsci-archive.pitt.edu/archive/00002681/01/lense.pdf}

%====================================================
\bibitem{painleve1}
Paul Painlev\'e, 
``La m\'ecanique classique et la th\'eorie de la relativit\'e\,", \\
C.~R.~Acad. Sci. (Paris) 173, 677--680(1921).

\bibitem{painleve2}
Paul Painlev\'e, \\
 ``La gravitation dans la m\'ecanique de Newton et dans la m\'ecanique d'Einstein'', \\
 C.~R.~Acad. Sci. (Paris) 173, 873--886(1921).
%\clearpage

%\clearpage
\bibitem{gullstrand}
Allvar Gullstrand,\\
``Allgemeine L\"osung des statischen Eink\"orperproblems in der Einsteinschen Gravitationstheorie",\\ 
 Arkiv f\"or Matematik, Astronomi och Fysik. 16 (8): 1--15 (1922).
 
 
 %=====================================================
 
 \bibitem{poisson}
K.~Martel and E.~Poisson,\\
``Regular coordinate systems for Schwarzschild and other spherical space-times'',\\
Am. J. Phys. \textbf{69} (2001), 476-480
\doi{10.1119/1.1336836}\\{}
[\arXiv{gr-qc/0001069} [gr-qc]].
%125 citations counted in INSPIRE as of 17 Jun 2020

\bibitem{Faraoni:2020}
V.~Faraoni and G.~Vachon,
``When Painlev\'e\textendash{}Gullstrand coordinates fail'',\\
Eur. Phys. J. C \textbf{80} (2020) no.8, 771
\doi{10.1140/epjc/s10052-020-8345-4}
[\arXiv{2006.10827} [gr-qc]].
%14 citations counted in INSPIRE as of 17 Feb 2022

\enlargethispage{20pt}
\bibitem{Boonserm:2017}
P.~Boonserm, T.~Ngampitipan and M.~Visser,
``Near-horizon geodesics for astrophysical and idealised black holes: Coordinate velocity and coordinate acceleration'',
Universe \textbf{4} (2018) no.6, 68
\doi{10.3390/universe4060068}
[\arXiv{1710.06139} [gr-qc]].
%1 citations counted in INSPIRE as of 17 Feb 2022

%\clearpage
%================================================
\bibitem{PGLT1}
Joshua Baines, Thomas Berry, Alex Simpson, and Matt Visser, \\
``Painleve-Gullstrand form of the Lense-Thirring spacetime'',
\\
Universe \textbf{7 \# 4} (2021) 105,
\doi{10.3390/universe704010}
[\arXiv{2006.14258} [gr-qc]].
%6 citations counted in INSPIRE as of 29 May 2021
%%%%%%

\bibitem{PGLT2}
Joshua Baines, Thomas Berry, Alex Simpson, and Matt Visser, \\
``Killing tensor and Carter constant for Painlev\'e--Gullstrand form of Lense--Thirring spacetime'',
Universe \textbf{7 \# 12} (2021) 473,
\doi{10.3390/universe7120473}
[\arXiv{2110.01814} [gr-qc]].

\bibitem{PGLT3}
Joshua Baines, Thomas Berry, Alex Simpson, and Matt Visser, \\
``Geodesics for Painlev\'e--Gullstrand form of Lense--Thirring spacetime'',\\
Universe \textbf{8 \#2} (2022) 115, 
\doi{10.3390/universe8020115}
[\arXiv{2112.05228} [gr-qc]].
%Universe 2022, 8(2), 115; https://doi.org/10.3390/universe8020115 

%=======================================
\bibitem{river}
A.~J.~Hamilton and J.~P.~Lisle,\\
``The river model of black holes'',\\
Am. J. Phys. \textbf{76} (2008), 519-532
\doi{10.1119/1.2830526}\\{}
[\arXiv{gr-qc/0411060} [gr-qc]].

%=====================================
 

 
%-------------------------------------------------------------
 \bibitem{Unit-lapse}
J.~Baines, T.~Berry, A.~Simpson and M.~Visser,
``Unit-lapse versions of the Kerr spacetime'',
Class. Quant. Grav. \textbf{38} (2021) no.5, 055001
\doi{10.1088/1361-6382/abd071}
[\arXiv{2008.03817} [gr-qc]].
%6 citations counted in INSPIRE as of 05 Jun 2021

%==================================
  
 \bibitem{Valiente-Kroon:2004a}
 J.~A.~Valiente Kroon,
``On the nonexistence of conformally flat slices in the Kerr and other stationary space-times'',
Phys. Rev. Lett. \textbf{92} (2004), 041101
\doi{10.1103/PhysRevLett.92.041101}
[\arXiv{gr-qc/0310048} [gr-qc]].
%36 citations counted in INSPIRE as of 29 May 2021

\bibitem{Valiente-Kroon:2004b}
J.~A.~Valiente Kroon,
``Asymptotic expansions of the Cotton-York tensor on slices of stationary space-times'',
Class. Quant. Grav. \textbf{21} (2004), 3237-3250
\doi{10.1088/0264-9381/21/13/009}
[\arXiv{gr-qc/0402033} [gr-qc]].
%29 citations counted in INSPIRE as of 30 May 2021
 
 %\clearpage
 \bibitem{Jaramillo:2007}
J.~L.~Jaramillo, J.~A.~Valiente Kroon and E.~Gourgoulhon,
``From geometry to numerics: Interdisciplinary aspects in mathematical and numerical relativity'',\\
Class. Quant. Grav. \textbf{25} (2008), 093001
\doi{10.1088/0264-9381/25/9/093001}
[\arXiv{0712.2332} [gr-qc]].
%18 citations counted in INSPIRE as of 30 May 2021
  
 \bibitem{Kerr-Darboux}
Joshua Baines, Thomas Berry, Alex Simpson, and Matt Visser, \\
``Darboux diagonalization of the spatial 3-metric in Kerr spacetime",\\
  Gen.Rel.Grav. \textbf{53} (2021) 1, 3 \doi{10.1007/s10714-020-02765-0}
  \\{} [\arXiv{2009.01397} [gr-qc]]

%==================================
  
\bibitem{Papadopoulos:2020}
G.~O.~Papadopoulos and K.~D.~Kokkotas,\\
``On Kerr black hole deformations admitting a Carter constant and an invariant criterion for the separability of the wave equation'',\\
Gen. Rel. Grav. \textbf{53} (2021) no.2, 21
\doi{10.1007/s10714-021-02795-2}
{}[\arXiv{2007.12125} [gr-qc]].
%0 citations counted in INSPIRE as of 15 Sep 2020

\bibitem{Papadopoulos:2018}
G.~O.~Papadopoulos and K.~D.~Kokkotas,\\
``Preserving Kerr symmetries in deformed spacetimes'',\\
Class. Quant. Grav. \textbf{35} (2018) no.18, 185014
\doi{10.1088/1361-6382/aad7f4}\\{} [\arXiv{1807.08594} [gr-qc]].
%12 citations counted in INSPIRE as of 15 Sep 2020

%\clearpage
\bibitem{Benenti:1979}
S. Benenti and M. Francaviglia,
``Remarks on Certain Separability Structures and Their Applications to General Relativity'', \\
General Relativity and Gravitation {\bf10} (1979) 79--92.


%====================================
%=================================
%====================================
%\clearpage
 %====================================
%=====================================
%======================================
%==================================
 
 \bibitem{Carballo-Rubio:2018}
R.~Carballo-Rubio, F.~Di Filippo, S.~Liberati and M.~Visser,
``Phenomenological aspects of black holes beyond general relativity'',
Phys. Rev. D \textbf{98} (2018) no.12, 124009
\doi{10.1103/PhysRevD.98.124009}
[\arXiv{1809.08238} [gr-qc]].
%75 citations counted in INSPIRE as of 05 Jun 2021

\bibitem{Carballo-Rubio:2019}
R.~Carballo-Rubio, F.~Di Filippo, S.~Liberati and M.~Visser,
``Geodesically complete black holes'',
Phys. Rev. D \textbf{101} (2020), 084047
\doi{10.1103/PhysRevD.101.084047}
[\arXiv{1911.11200} [gr-qc]].
%49 citations counted in INSPIRE as of 11 Feb 2022

\bibitem{Carballo-Rubio:2021}
R.~Carballo-Rubio, F.~Di Filippo, S.~Liberati and M.~Visser,
``Geodesically complete black holes in Lorentz-violating gravity'',
JHEP 2022 (in press).
[\arXiv{2111.03113} [gr-qc]].
%1 citations counted in INSPIRE as of 11 Feb 2022

\bibitem{Pandora}
R.~Carballo-Rubio, F.~Di Filippo, S.~Liberati and M.~Visser,
``Opening the Pandora\textquoteright{}s box at the core of black holes'',
Class. Quant. Grav. \textbf{37} (2020) no.14, 14
\doi{10.1088/1361-6382/ab8141}
[\arXiv{1908.03261} [gr-qc]].
%39 citations counted in INSPIRE as of 11 Feb 2022

\bibitem{Simpson:2021a}
A.~Simpson and M.~Visser,
``The eye of the storm: A regular Kerr black hole'',
JCAP (in press), 
[\arXiv{2111.12329} [gr-qc]].
%5 citations counted in INSPIRE as of 11 Feb 2022

\bibitem{Simpson:2021b}
A.~Simpson and M.~Visser,
``Astrophysically viable Kerr-like spacetime -- into the eye of the storm'',
[\arXiv{2112.04647} [gr-qc]].
%0 citations counted in INSPIRE as of 11 Feb 2022

\bibitem{Visser:2009a}
M.~Visser, C.~Barcel\'o, S.~Liberati and S.~Sonego,\\
``Small, dark, and heavy: But is it a black hole?'',\\
PoS \textbf{BHGRS} (2008), 010
\doi{10.22323/1.075.0010}
[\arXiv{0902.0346} [gr-qc]].
%54 citations counted in INSPIRE as of 05 Jun 2021

\bibitem{Visser:2009b}
M.~Visser,
``Black holes in general relativity'',
PoS \textbf{BHGRS} (2008), 001
\doi{10.22323/1.075.0001}
[\arXiv{0901.4365} [gr-qc]].
%38 citations counted in INSPIRE as of 05 Jun 2021

%===================================

\bibitem{bb-Kerr}
J.~Mazza, E.~Franzin and S.~Liberati,
``A novel family of rotating black hole mimickers'',
JCAP \textbf{04} (2021), 082
\doi{10.1088/1475-7516/2021/04/082}
[\arXiv{2102.01105} [gr-qc]].
%10 citations counted in INSPIRE as of 05 Jun 2021

\bibitem{bb-KN}
E.~Franzin, S.~Liberati, J.~Mazza, A.~Simpson and M.~Visser,
``Charged black-bounce spacetimes'',
JCAP \textbf{07} (2021), 036
\doi{10.1088/1475-7516/2021/07/036}
[\arXiv{2104.11376} [gr-qc]].
%0 citations counted in INSPIRE as of 30 May 2021

\bibitem{DeLorenzo:2014}
T.~De Lorenzo, C.~Pacilio, C.~Rovelli and S.~Speziale,
``On the Effective Metric of a Planck Star'',
Gen. Rel. Grav. \textbf{47} (2015) no.4, 41
\doi{10.1007/s10714-015-1882-8}
[\arXiv{1412.6015} [gr-qc]].
%81 citations counted in INSPIRE as of 11 Feb 2022

\bibitem{Hayward:2005}
S.~A.~Hayward,
``Formation and evaporation of regular black holes'',
Phys. Rev. Lett. \textbf{96} (2006), 031103
\doi{10.1103/PhysRevLett.96.031103}
[\arXiv{gr-qc/0506126} [gr-qc]].
%624 citations counted in INSPIRE as of 11 Feb 2022

\bibitem{Bronnikov:2000}
K.~A.~Bronnikov,
``Regular magnetic black holes and monopoles from nonlinear electrodynamics'',
Phys. Rev. D \textbf{63} (2001), 044005
\doi{10.1103/PhysRevD.63.044005}
[\arXiv{gr-qc/0006014} [gr-qc]].
%416 citations counted in INSPIRE as of 11 Feb 2022

\bibitem{Bronnikov:2005}
K.~A.~Bronnikov and J.~C.~Fabris,
``Regular phantom black holes'',
Phys. Rev. Lett. \textbf{96} (2006), 251101
\doi{10.1103/PhysRevLett.96.251101}
[\arXiv{gr-qc/0511109} [gr-qc]].
%244 citations counted in INSPIRE as of 11 Feb 2022

\bibitem{Johannsen:2011}
T.~Johannsen and D.~Psaltis,
``A Metric for Rapidly Spinning Black Holes Suitable for Strong-Field Tests of the No-Hair Theorem'',
Phys. Rev. D \textbf{83} (2011), 124015
\doi{10.1103/PhysRevD.83.124015}
[\arXiv{1105.3191} [gr-qc]].
%220 citations counted in INSPIRE as of 11 Feb 2022

\bibitem{Bardeen:1968}
J.~M.~Bardeen, ``Non-singular general relativistic gravitational collapse'',
Abstracts of the 5th international conference on gravitation 
and the theory of relativity (GR5), 
eds. V.~A.~Fock \emph{et al.} 
(Tbilisi University Press, Tblisi, Georgia,  former USSR, 1968), 
pages 174--175.



%=====================================
%=====================================
%------------------------------------
\bigskip
\hrule\hrule\hrule
\bigskip
%------------------------------------

%===================================
%=================================
%================================
%==========================
%==========================
%==========================
%=================================
\end{thebibliography}
\end{document}